\documentclass[prl,longbibliography,preprint]{revtex4-1}
\usepackage{graphicx}
\usepackage{amsmath}
\usepackage{amssymb}
\usepackage{txfonts}
\usepackage{comment}
\everymath{\displaystyle}
\begin{document}

\title{A Universal Molecular-Kinetic Scaling Relation for Slip of a Simple Fluid at a Solid Boundary} %Title of paper
\author{Gerald J. Wang and Nicolas G. Hadjiconstantinou}

\affiliation{Department of Mechanical Engineering, Massachusetts Institute of Technology\\Cambridge, MA 02139, USA}

%\date{\today}

\begin{abstract}
Using the observation that slip in simple fluids at low and moderate shear rates is a thermally activated process driven by the shear stress in the fluid close to the solid boundary, we develop a molecular-kinetic model for simple fluid slip at solid boundaries. The proposed model, which is in the form of a universal scaling relation that connects slip and shear rate, reduces to the well known Navier-slip condition under low shear conditions, providing a direct connection between molecular parameters and the slip length. Molecular-dynamics simulations are in very good agreement with the predicted dependence of slip on system parameters, including the temperature and fluid-solid interaction strength. Connections between our model and previous work, as well as simulation and experimental results, are explored and discussed.
\end{abstract}

\pacs{}% insert suggested PACS numbers in braces on next line

\keywords{Slip, nanoconfined fluids, interfacial phenomena, nanoscale fluid-solid interactions}

\maketitle %\maketitle must follow title, authors, abstract and \pacs

\section{Introduction} Fluids under nanoscale confinement can exhibit a number of remarkable transport properties, including anomalous flow rates \cite{Hummer, Majumder, ThomasMcGaugheySubcontinuum}, diffusion \cite{FirstLayerDiffusivity}, and heat transfer \cite{FLGKapitza}. Nanofluidic engineering exploits these phenomena for the development of novel materials for clean water \cite{KarnikDesalReview} and energy \cite{TengfeiLuoEnergyInterfaces} among other applications. Modeling the dynamics of nanoconfined fluids requires a detailed understanding of the effect of the fluid-solid interface on transport in these conditions \cite{FluidsInCNTs,ThompsonRobbinsStickSlip,StrainEngineeringFLG,ThompsonTroian}. 

Slip at the fluid-solid interface is, perhaps, the most ubiquitous of these phenomena and has received considerable attention (see, for example, \cite{ThompsonTroian, VinogradovaSlip, deGennesSlip, FirstLayerSlip, Lichter_vdFK, PriezjevTroian, ShuTeoChanReview,PriezjevSlip,RateDependentSlipBoundary}). In the case of dilute gas systems, the functional form of the slip relation as well as slip coefficients can be calculated via asymptotic expansions of the Boltzmann equation \cite{Sone2007, pof2006, JP2016}. In dense liquids such analytical treatments are not possible; however, strong empirical evidence exists that the slip relation is of the same form as the dilute case, namely
\begin{equation}
u_s=\beta\frac{\partial u}{\partial \eta}\bigg|_{\text{b}}
\label{Navier}
\end{equation}
which is referred to as the Navier slip condition, after Navier \cite{NavierSlip} who first proposed it. In this expression, $u_s$ is the slip velocity (difference between the fluid velocity at the boundary  and the boundary velocity), $u$ is the flow velocity in the direction parallel to the boundary, and $\eta$ is the wall normal in the direction pointing into the fluid; the subscript ``b"  denotes the boundary location. 

Most research in the dense-fluid arena has thus focused on investigating the properties of the slip length $\beta$. Of particular note is the work of Thompson and Troian \cite{ThompsonTroian}, which showed that molecular-dynamics (MD) data for the slip length  could be described well by an expression of the form $\beta=\beta_0(1-\dot\gamma/\dot\gamma_c)^{-1/2}$ (where $\beta_0$ is the slip length at low shear rates, $\dot \gamma$ denotes the shear rate, and $\dot \gamma_c$ is a constant that depends on the fluid as well as the fluid-solid interaction details), suggesting that the slip length obeys some form of critical dynamics. Other groups have made use of Green-Kubo analyses to develop models that relate the slip length to the solid-liquid interaction potential corrugations, both at the atomistic scale \cite{FirstLayerSlip, BocquetBarratGKFriction} and the roughness scale  \cite{PriezjevTroian}. Several authors \cite{BlakeHaynes, LichterSlipRateProcess, MartiniLichterJFM} have proposed that slip exhibits many of the hallmarks of a thermally activated process, at least for simple fluids in contact with atomically smooth boundaries. 

Despite this considerable progress, complete and predictive models of fluid slip based on ab-initio (molecular) considerations have yet to be fully developed. The goal of the present work is to present a physically motivated model for slip at the interface between a simple fluid and a molecularly smooth solid that is able to unify existing work and explain our, as well as previous, simulation results.

\section{Formulation} The model proposed here is based on the observation that fluid slip on a molecularly smooth wall at low and moderate shear rates can be modeled as a thermally activated process \cite{LichterSlipRateProcess, MartiniLichterJFM} and can thus be quantitatively described using an extension of the Eyring theory of reaction rates \cite{EyringReactionRates,Hirschfelder}. For a discussion of this theory and its connection to transition state theory (TST), see \cite{HanggiKramers}. Provided the conditions for an activated process are met, rate theory can be used to relate the drift velocity of molecules under the influence of a driving force to the rate of hopping over the potential barrier generated by nearby molecules. Eyring used such an approach to develop a theory for the viscosities of dense fluids \cite{Hirschfelder}, while Blake and coworkers pioneered the use of this concept to describe slip and utilized it to model  contact-line motion as a molecular-kinetic process \cite{BlakeHaynes}. The model by Blake and Haynes (in particular, \cite{BlakeHaynes}) has been widely accepted in the contact-line literature and has been found to be in good agreement with experimental data (see \cite{Eggers,BlakeReview} for discussions) as well as MD simulations of contact-line motion \cite{Ren2007,ItalianPaper}.

Following the exposition by Wyart and deGennes \cite{Wyart_and_deGennes}, the drift velocity, $u_\text{d}$, can be written as the difference between the forward and backward hopping rates $u_\text{d}=l_j(\kappa^+-\kappa^-)$ with $$\kappa^\pm=\tau_0^{-1}\exp\bigg(-\frac{V\mp \frac{1}{2}f_\text{d} l_j^2}{k_BT}\bigg)$$ leading to
\begin{equation}
u_\text{d}=\frac{2l_j }{\tau_0}\exp\bigg(-\frac{V}{k_BT}\bigg)\sinh\bigg(\frac{f_\text{d}l_j^2}{2k_BT}\bigg)
\end{equation}
where $l_j$ is the jump length, $\tau_0$ is the jump time scale, $V$ is the potential barrier associated with the jump, $k_\text{B}$ is Boltzmann's constant, and $f_\text{d}$ is the force per unit length acting on the fluid molecules  causing this drift.

To make further progress, we make two observations. First, in the case of slip, the force on the molecules at the fluid-solid boundary responsible for the drift is the shear stress in the fluid at this location, $\mu (\partial u/\partial \eta)|_\text{b}$, where $\mu$ denotes the viscosity. Assuming that within a few atomic diameters from the boundary an ``inner" description exists in which molecular effects dominate, and noting that the slip boundary condition is associated with the outer (Navier-Stokes) region description, we interpret the above quantity as taken at the interface of the outer and inner regions, that is, in a region where the Navier-Stokes description is still valid. In other words, $\mu$ corresponds to the bulk value of the viscosity, while $(\partial u/\partial \eta)|_\text{b}$ corresponds to the velocity gradient a few atomic diameters away from the boundary where layering effects do not affect the flow field significantly.

Our second observation is that, in its most general form, the potential $V$ represents the overall potential landscape and thus for a fluid molecule at the fluid-solid interface it includes both the fluid-solid ($V_{s}$) as well as fluid-fluid ($V_{f}$) interactions. Assuming additivity of the potential contributions, which is certainly true for our simulations, $\exp(-V/k_BT)$ can be factored into a term containing the fluid-solid interaction and a term containing the fluid-fluid interaction, which can be absorbed into the time scale by defining $\hat{\tau}_0=\tau_0 \exp(V_{f}/k_BT)$. As shown below, this allows us to explicitly highlight the effect of fluid-solid interaction.  

Writing $f_\text{d} l_j=\mu \Sigma_\text{FL}^{-1}(\partial u/\partial \eta)|_\text{b} $, where $\Sigma_\text{FL}$ denotes the areal density of fluid molecules at the fluid-solid interface (number of molecules in the first fluid layer at the fluid-wall interface per unit interface area), we obtain the following expression for the slip velocity
\begin{equation}
\label{the_big_kahuna}
u_s=\frac{2l_j }{\hat{\tau}_0}\exp\bigg(-\frac{\alpha\varepsilon}{k_BT}\bigg)\sinh\bigg(\frac{\mu l_j}{2\Sigma_\text{FL} k_BT}\dot\gamma\bigg)
\end{equation}
where, as a reminder, $\hat{\tau}_0$ now includes the contribution of fluid environment on the potential barrier. In this expression, we have factored the overall fluid-solid interaction energy $V_{s}$ into $\alpha\varepsilon$, where $\varepsilon$ is the energy scale for fluid-solid interactions and $\alpha$  represents the potential energy of each fluid atom in the first fluid layer due to its interaction with all of the solid atoms, expressed in units of $\varepsilon$. In other words, $\alpha$  is the scaled potential interaction energy of  a single fluid atom in the first fluid layer summed (or, in the mean-field sense, integrated) over all solid atoms. The properties and characteristic values of the areal density, $\Sigma_\text{FL}$, and the related volumetric density of the first fluid layer at the fluid-solid interface, $\rho_\text{FL}=\Sigma_\text{FL}/h_\text{FL}$ (where $h_\text{FL}$ denotes the width of the first layer), have been extensively studied in a recent publication by the authors \cite{FirstLayerDensity}. 

It immediately follows from (\ref{the_big_kahuna}) that in the small-shear-rate limit,
$\dot \gamma\ll 2\Sigma_\text{FL} k_BT/(l_j\mu)$,
 Eqn. (\ref{the_big_kahuna}) linearizes to the Navier-slip condition \eqref{Navier}, with 
 \begin{equation}
 \beta = \frac{\mu l_j^2}{\Sigma_\text{FL} k_BT\hat{\tau}_0} \exp\bigg(-\frac{\alpha \varepsilon}{k_BT} \bigg)
 \label{linear}
 \end{equation}
 It is worth observing that, contrary to the continuum approach where $\beta$ is a parameter whose value needs to be supplied as part of the problem specification, the molecular-kinetic approach provides a direct connection between the slip length and the governing molecular parameters. As a consequence, with very detailed micro-mechanical information, Eqn. (\ref{linear}) could in principle  be used to directly predict the slip length from first principles.

\section{Validation} 
To assess these ideas, we performed non-equilibrium MD simulations of plane-Couette flow, described in detail in the Appendix. Our results and subsequent discussion will be expressed in terms of standard LJ non-dimensional quantities \cite{Allen}, namely, $\sigma$ for length, $\varepsilon_f$ for energy, and $\tau\equiv(m\sigma^2/\varepsilon_f)^{1/2}$ for time. In order to verify each of the dependences in Eqn. (\ref{the_big_kahuna}), we measure the slip velocity as we systematically vary the shear rate, the temperature, and the fluid-solid interaction strength.

\begin{figure}[h]
\includegraphics[width=\textwidth]{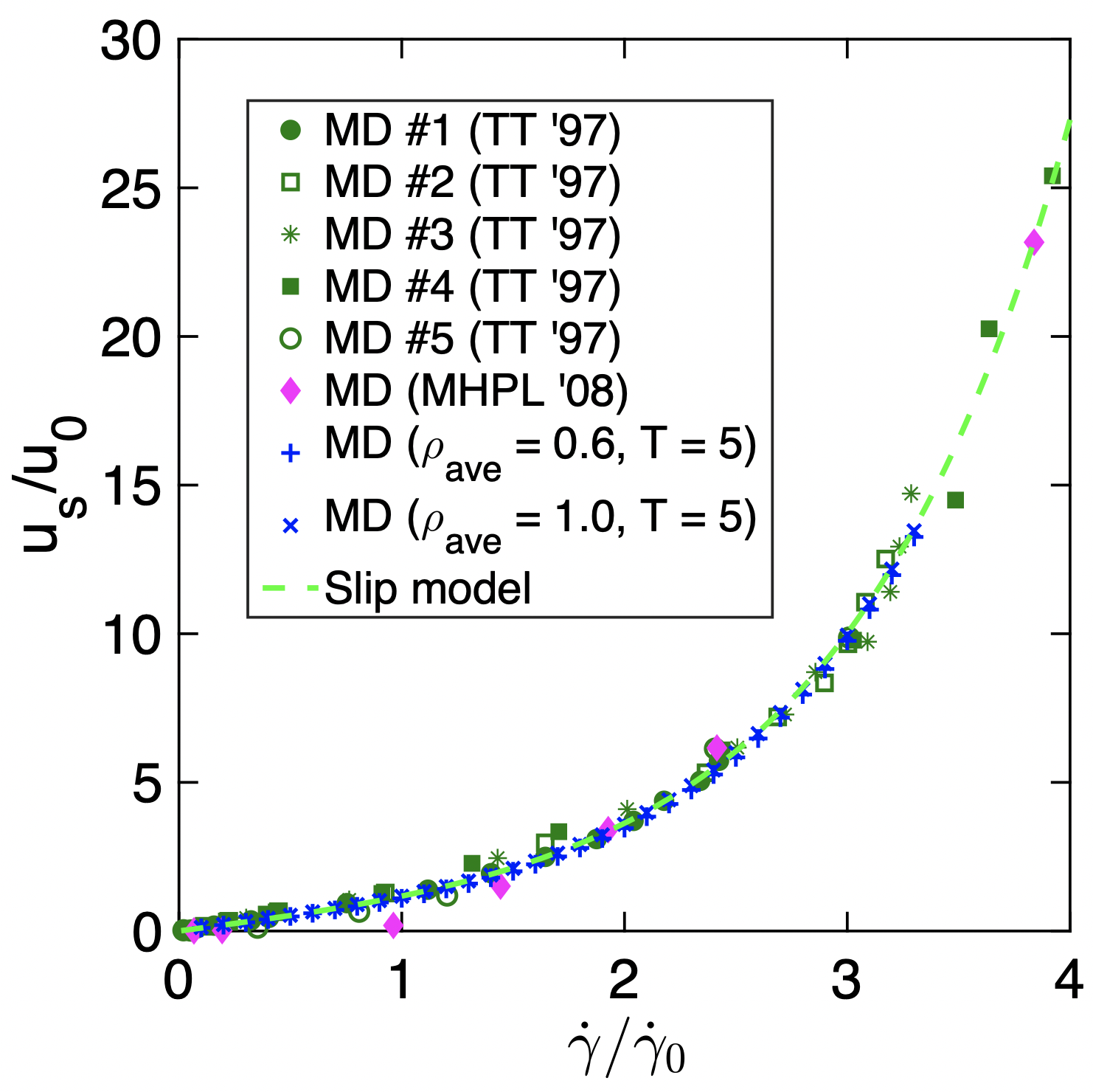}
\caption{\label{slip_vs_shear_MD} Scaled slip velocity as a function of scaled shear as measured in MD simulations. The scaling predicted by Eqn. (\ref{the_big_kahuna}) is shown as the dashed green line (Slip model). MD data transcribed from \cite{ThompsonTroian} (TT '97) and \cite{MartiniSlip} (MHPL '08) and scaled according to Eqn. (\ref{the_big_kahuna}) are shown in green and pink; in-house MD results are shown in blue.}
\end{figure}

\begin{figure}[h]
\includegraphics[width=\textwidth]{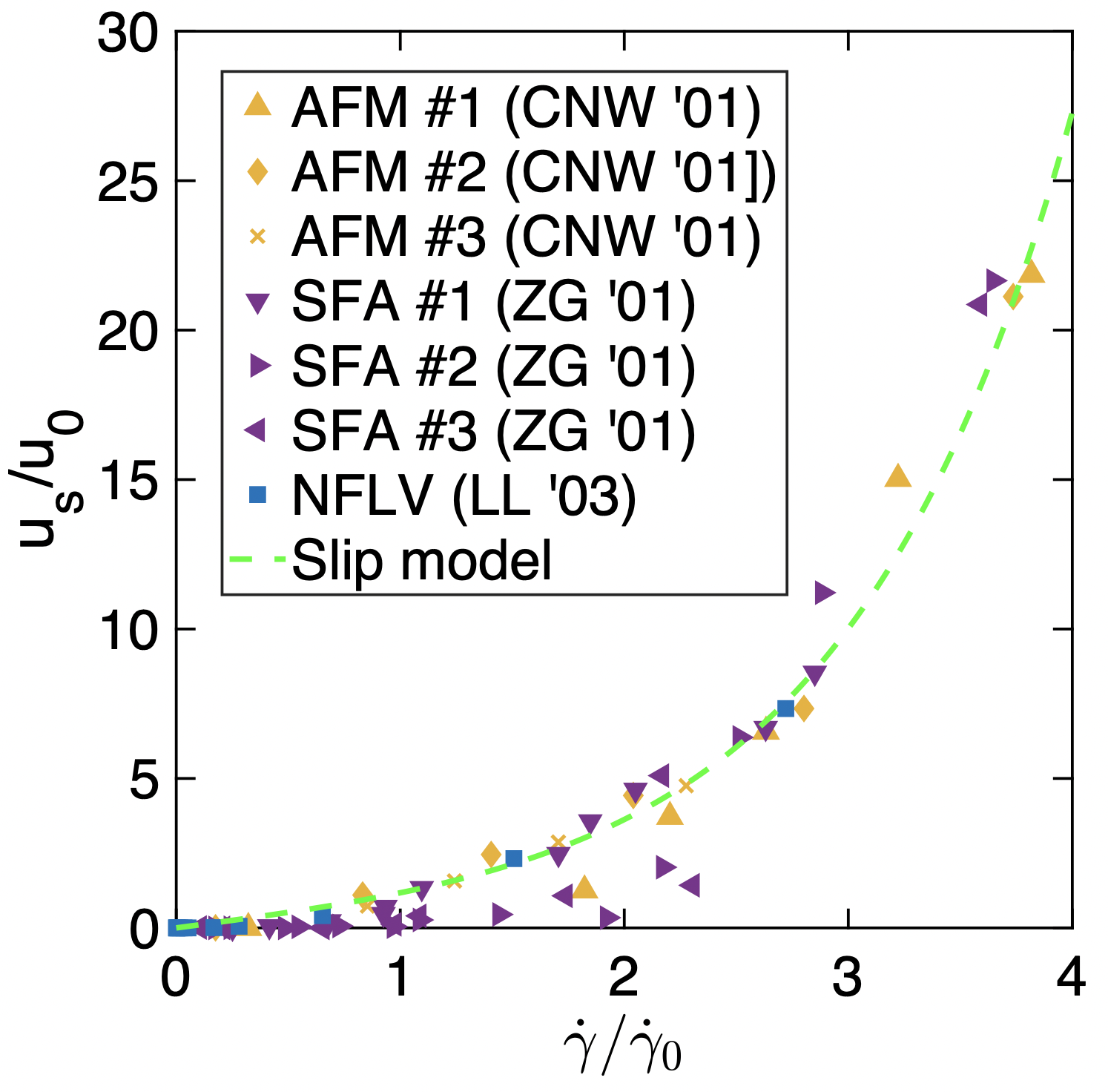}
\caption{\label{slip_vs_shear_experiments} Scaled slip velocity as a function of scaled shear for experiments performed via atomic force microscopy \cite{CraigNetoWilliams} (CNW '01), surface-force apparatus \cite{ZhuGranickSlip} (ZG '01), and near-field laser velocimetry \cite{LegerSlip} (LL '03). The scaling predicted by Eqn. (\ref{the_big_kahuna}) is shown as the dashed green line (Slip model).}
\end{figure}

We begin by studying the dependence on shear rate. Figure \ref{slip_vs_shear_MD} shows a comparison between MD simulation results and the prediction of Eqn. (\ref{the_big_kahuna}) scaled in the form 
\begin{equation}
\label{shear_rate_scaling}
u_s=u_0 \sinh (\dot\gamma/\dot\gamma_0).
\end{equation} To test this scaling, we generated datasets in the above-described geometry in which the shear rate was varied, while all other parameters were held constant. To augment these datasets, we also collected all rigid-wall MD simulation data from \cite{ThompsonTroian} and \cite{MartiniSlip} and scaled those according to (\ref{shear_rate_scaling}). In plotting the scaled data, the constants $u_0\equiv \frac{2l_j }{\hat{\tau}_0}\exp\bigg(-\frac{\alpha\varepsilon}{k_BT}\bigg)$ and $\dot\gamma_0\equiv \frac{2\Sigma_\text{FL} k_BT}{\mu l_j}$ were determined for each dataset by means of a non-linear least-squares fit to \eqref{shear_rate_scaling}. For in-house MD simulations (shown in blue), the vertical size of the symbols reflects the characteristic scale of uncertainty in the corresponding slip velocity measurements. In particular, the symbol height is equal to the width of the 95\% confidence interval on the slip velocity, as determined via a linear fit to the velocity profile in the bulk region (as described in the Appendix). The figure shows that Eqn. \eqref{shear_rate_scaling} is able to accurately describe all datasets, including those reported by Thompson and Troian, which were shown to be fitted well by the expression $\beta=\beta_0(1-\dot\gamma/\dot\gamma_c)^{-1/2}$ \cite{ThompsonTroian}. In other words, although both (\ref{shear_rate_scaling}) and the model by Thompson and Troian fit the data of Figure \ref{slip_vs_shear_MD} with comparable accuracy in view of the uncertainty in the data, compared to the latter, expression (\ref{shear_rate_scaling}) has the benefit of being more physically motivated, both due to its clear connection to a physical model of slip and the fact that it predicts finite slip at all finite shear rates.  In fact, \eqref{shear_rate_scaling} is also able to describe well the slip velocities measured in a wide range \cite{CraigNetoWilliams, ZhuGranickSlip, LegerSlip} of experiments (see Fig. \ref{slip_vs_shear_experiments}) involving aqueous solutions, alkanes, and polymeric fluids. In both figures, we are able to observe a regime of shear rates consistent with the Navier slip relation (for $\dot\gamma/\dot\gamma_0\lesssim 1$, $u_s\propto \dot{\gamma}$), beyond which the slip velocity rises dramatically with shear rate. 

Using values of the viscosity $\mu$ from \cite{LJ_Viscosity} and values of the areal density of interfacial fluid $\Sigma_\text{FL}$ from \cite{FirstLayerDensity}, we are able to infer the values for the molecular-kinetic parameters in \eqref{the_big_kahuna}, namely, $l_j$ and $\hat{\tau}_0 \exp\bigg(\frac{\alpha\varepsilon}{k_BT}\bigg)$ for our in-house MD data. These values, along with their associated 95\% confidence interval, are presented in Table I. We observe that both parameters are of molecular scale as expected. In fact, $l_j$ lies between the graphene interatomic spacing ($a=1.42\AA$) and the fluid-fluid characteristic spacing ($\sigma=3.15 \AA$), demonstrating consistency with the proposed molecular mechanism. Here we note that the slight increase of $l_j$ with $\rho_\text{ave}$ suggests that the effect of fluid-fluid interaction becomes more important as the density is increased, as expected. Along the same lines, we also note the sensitivity of the product $\hat{\tau}_0 \exp\bigg(\frac{\alpha\varepsilon}{k_BT}\bigg)$  to $\rho_\text{ave}$; this can be understood by noting that $V_f$ is expected to increase with $\rho_\text{ave}$, again demonstrating consistency with the proposed molecular mechanism. Further analysis of the complete term $\hat{\tau}_0 \exp\bigg(\frac{\alpha\varepsilon}{k_BT}\bigg)$, including development of approaches for separating the contributions of its constituents, will be undertaken in the future.

\begin{table}[!htbp]
\begin{tabular}{c||c|c}
\label{tab:parameter_values}
$\rho_\text{ave}$& $l_j$ [$\AA$] & $\hat{\tau}_0 \exp\bigg(\frac{\alpha\varepsilon}{k_BT}\bigg)$ {[}ns{]} \\
\hline\hline
0.6                        & $1.6\pm 0.29$                    & $0.034 \pm 0.009$                         \\
1.0                        & $2.4\pm 0.35$                    & $0.92 \pm 0.51$                         
\end{tabular}
\caption{Parameters for characteristic length and time scales in \eqref{the_big_kahuna}, for each of the densities in Fig. \ref{slip_vs_shear_MD}. }
\end{table}

We now investigate the other major factors affecting slip, namely, the temperature and the fluid-solid interaction strength. Having verified the non-linear behavior with shear rate in the above section, the following comparisons will be performed for low shear rates for which an explicit expression for the slip length exists (see (\ref{linear})). 

\begin{figure}[h]
\includegraphics[width=\textwidth]{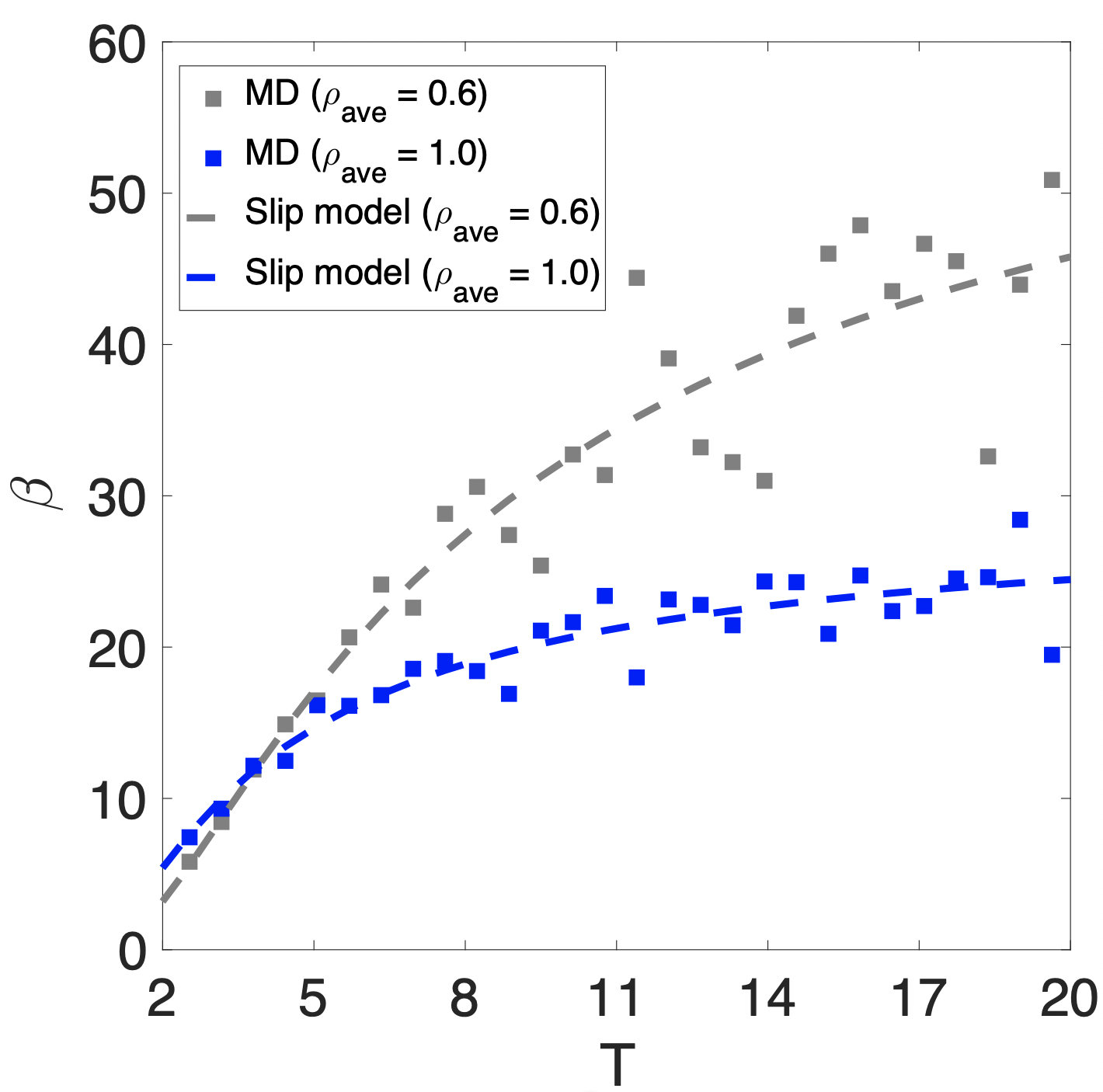}
\caption{\label{slip_vs_temperature} Slip length as a function of temperature for high- and low-density fluids ($u_w=0.25$), where the change in non-dimensional temperature is effected by changes in the dimensional temperature ($\varepsilon$ is held constant at 1). For both densities, the results from MD simulation show strong agreement with the dependence in \eqref{temperature_fit} (Slip model).}
\end{figure}

Figure \ref{slip_vs_temperature} shows the temperature dependence of the slip length at a fixed fluid-solid interaction strength $\varepsilon=1$ and boundary speed $u_w=0.25$, for two fluid densities. We find that the slip length is fitted well ($R^2 \geq 0.89$) by the form 
\begin{equation}
\label{temperature_fit}
\beta= c_1\exp\bigg(\frac{c_2}{T}\bigg)
\end{equation}
where $c_1$ and $c_2$ are fitting constants. This form represents the dominant temperature dependence in \eqref{linear} arising from term $\exp(-V/k_BT)$. Additional contributions to the temperature dependence can be accounted for by noting that (a) for a Lennard-Jones fluid, $\mu\sim T^\zeta$ with $\zeta \approx 1$ when $\rho\gtrsim 0.6$ and $T\gtrsim 2$ \cite{LJTransportCoeff} and (b) the leading-order temperature dependence of $\Sigma_\text{FL}$ was shown in \cite{FirstLayerDensity} to be of the form $\Sigma_\text{FL}\sim a+b \text{ Wa}$, where $\text{Wa}\propto \varepsilon/(k_BT)$ is the Wall number introduced in \cite{FirstLayerDensity} and $a$ and $b$ are density-dependent constants introduced and discussed in Section II C of the same reference. Modifying expression (\ref{temperature_fit}) to  $\beta=c_1\exp(c_2/T) T^\gamma(a+b'/T)^{-1}$, where $T^\gamma$ represents any residual temperature dependence  (e.g. $\tau_0$ or due to the difference between $\zeta$ and 1)  and using values of $a$ and $b'$ as calculated from independent MD simulations in \cite{FirstLayerDensity}, has a negligible effect on the fit quality ($R^2$ value increase of 1\% or less).

\begin{figure}[h]
\includegraphics[width=\textwidth]{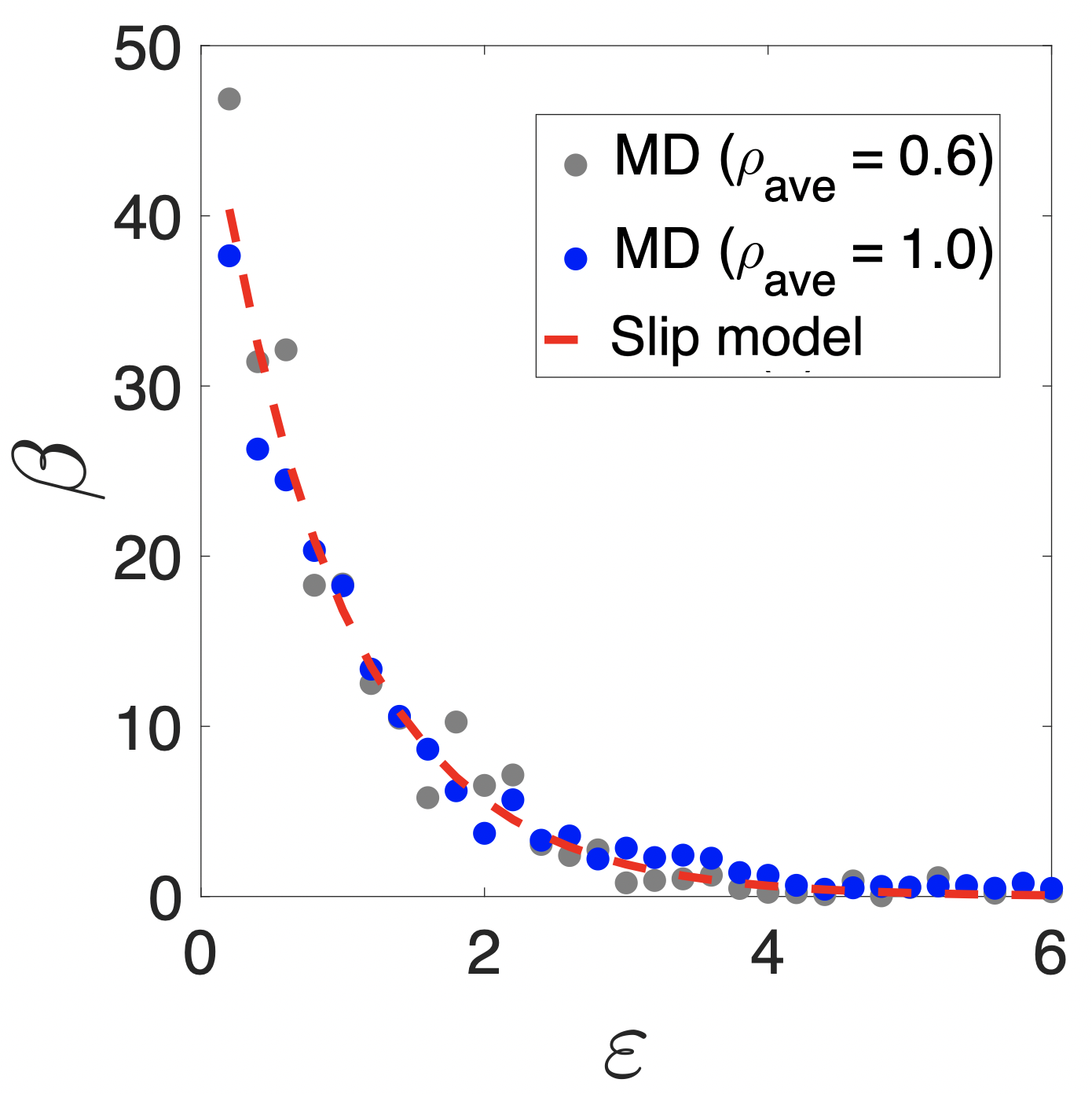}
\caption{\label{slip_vs_epsilon} Slip length as a function of fluid-solid interaction strength for high- and low-density fluids ($u_w=0.125$). For both densities, the results from MD simulation show strong agreement with the exponential decay given by \eqref{epsilon_fit} (Slip model).}
\end{figure}

Figure \ref{slip_vs_epsilon} shows the dependence of the slip length on $\varepsilon$ at a fixed temperature $T=5$ and boundary speed $u_w=0.125$, for two fluid densities. We find that the form
\begin{equation}
\label{epsilon_fit}
\beta = c_3 \exp (-c_4 \varepsilon)
\end{equation}
where $c_3$ and $c_4$ are fitting constants, results in a very strong fit ($R^2\geq 0.91$).  This form represents the dominant dependence of $\beta$ on $\varepsilon$ in \eqref{linear}.  Including the dependence of $\Sigma_{FL}$ on $\varepsilon$ by modifying expression (\ref{epsilon_fit}) to  $\beta=c_3\exp(-c_4 \varepsilon) (a+b''\varepsilon )^{-1}$ and using values of $a$ and $b''$ as calculated  in \cite{FirstLayerDensity} has a negligible effect on the fit quality ($R^2$ value increases from 0.971 to 0.973). We also note that for the set of simulations depicted in Fig. \ref{slip_vs_epsilon} the slip length is essentially independent of the liquid density  (in fact, both data sets are jointly well fitted by the same exponential decay). This is consistent with the results of Fig. \ref{slip_vs_temperature} which shows that the slip length is nearly independent of density for $T\lesssim 5$.  As a final note on the consistency of the two fits ((\ref{temperature_fit}) and (\ref{epsilon_fit})), we point out that at the condition $T=5$ and $\varepsilon=1$, both predict slip length values of approximately 16, for both fluid densities.

\section{Connection to other studies of slip}
The present model is, in general,  in good agreement with  earlier theoretical  and experimental work on liquid slip. For example, Eqn. (\ref{the_big_kahuna}) predicts that at fixed temperature, in the small-shear-rate limit, slip is linear in the bulk viscosity. This is in good agreement with the Green-Kubo analysis by Barrat and Bocquet \cite{FirstLayerSlip, BocquetBarratGKFriction}, who found the same scaling; this observation also agrees qualitatively with the predictions of the variable-density Frenkel-Kontorova model \cite{Lichter_vdFK}. The strength of this result is even clearer after careful comparison with experiments that probe the effect of varying both the viscosity and the shear rate \cite{CraigNetoWilliams} across the transition into the non-linear regime. The data of Craig \emph{et al.}, included in Fig. \ref{slip_vs_shear_experiments}, show both a transition to the non-linear regime as the shear rate is increased at fixed viscosity, but also that the controlling factor for this transition is the product of viscosity and shear rate. Their data shows quantitative agreement (within 14\% error) with the predictions of \eqref{the_big_kahuna} (see Fig. \ref{slip_vs_shear_experiments}). 

Equation (\ref{the_big_kahuna}) is also in good qualitative agreement with earlier work focused on the relationship between slip and wettability. Both MD simulations \cite{ThompsonTroian, BarratBocquet_LargeSlipEffect} and experiments \cite{Baudry_ExperimentalSlip, Cottin-Bizonne_ExperimentalSlip} have found that slip tends to increase as the fluid becomes less wetting (as the energy scale $\varepsilon$ of fluid-solid interaction decreases). More specifically, the MD simulations in \cite{BarratBocquet_LargeSlipEffect} strongly suggest that $d\beta/d\varepsilon<0$ and $d^2\beta/d\varepsilon^2>0$, i.e. slip decays as a convex function of $\varepsilon$, which is reflected by Eqn. (\ref{the_big_kahuna}). We also note that under some assumptions, Green-Kubo theory \cite{FirstLayerSlip} predicts $\beta \propto \varepsilon^{-2}$ (for fixed first layer density and first layer structure factor) which is in qualitative agreement with the above findings; this behavior was also observed in MD simulations of  patterned wettability \cite{PriezjevTroianPatternedWettability} in the cases where the slip is determined by the interaction between the first fluid layer and the wall. Here, we also note that in the limit $\varepsilon \rightarrow 0$, where the barrier height associated with the fluid-solid interaction becomes small, we expect (\ref{the_big_kahuna}) to no longer hold. 

As explained above, the presence of $\Sigma_{FL}$ in equation \eqref{the_big_kahuna} provides a direct connection between slip and fluid layering at the fluid-solid interface; this connection was first studied by Barrat and Bocquet \cite{FirstLayerSlip} in atomically smooth settings using Green-Kubo theory. The latter study has shown that, under certain assumptions,  and under fixed temperature, fluid-solid interaction and first layer structure factor as well as diffusion coefficient in the direction parallel to the fluid-solid interface in the first fluid layer, $\beta \propto 1/\rho_\text{FL}$. Careful MD simulations \cite{RateDependentSlipBoundary} have verified that the  slip length and $\rho_{FL}$ are inversely related, but with a slightly different exponent, namely $\beta\sim \rho_\text{FL}^{-1.44}$. These results are in general agreement with the prediction of Eqn. (\ref{linear}); the latter can be seen more clearly by noting that $\Sigma_\text{FL}=\rho_\text{FL}h_\text{FL}$ and that the dependence of $h_\text{FL}$ on density is weak \cite{FirstLayerDensity}; it also needs to be noted that the remaining  independent variables appearing in our expression are not exactly the same as the expression of Barrat and Bocquet.  The agreement between Green-Kubo theory and MD simulations was also studied in the presence of physical surface corrugations (wall roughness) in \cite{PriezjevTroian};  this is an important consideration that merits further examination from the rate-process perspective in the future.

\section{Conclusions and Outlook} We have proposed and validated a general scaling relation describing slip of simple fluids at smooth solid boundaries. The model builds upon the observation that slip is a thermally activated process, proposed by others \cite{BlakeHaynes} and more firmly established by recent careful studies \cite{LichterSlipRateProcess}. The proposed scaling relation is found to be in excellent agreement with MD simulations as well as experimental data; the latter even extends to moderately complex fluids. In other words, the proposed model provides further evidence that slip at low and moderate shear rates can be modeled as a rate process but also provides an explicit expression for predicting  slip in terms of molecular system parameters. Although the power-law relation proposed in \cite{ThompsonTroian} also exhibits good agreement with MD simulations, Eqn. \eqref{the_big_kahuna} has the advantage of being associated with a clear physical model of the slip process and does not suffer from a slip-length divergence at finite shear rates. 

The work by Martini et al. \cite{MartiniSlip} suggests that in the high rate limit, slip is no longer a rate process. In this work we have limited our MD simulations to low and moderate shear rates given by  $\dot\gamma\tau \lesssim 0.07$, such that, in addition to low-to-moderate shear rates, we ensure $\mu\neq\mu(\dot\gamma)$. It is worth emphasizing that, from a practical perspective, this low to moderate shear-rate regime covers virtually all nanofluidic engineering applications (for channels of nanoscale dimensions and typical fluids, this condition corresponds to flow velocities $\lesssim \mathcal O(10^2)$ m/s). On the other hand, we note that  at high shear rates Eqn. \eqref{the_big_kahuna} predicts a plateau for the slip length (shear thinning in the large-$\dot\gamma$ limit typically results in a behavior of the type  $\mu\propto \dot\gamma^{-1}$ for simple fluids \cite{HeyesShearThinning}), in qualitative agreement with the predictions of the Frenkel-Kontorova model and MD simulations \cite{MartiniSlip}. A quantitative investigation of the high-shear-rate limit will be the subject of future work. 

\section{Acknowledgements}
The authors are grateful for support from the DOE CSGF (Contract No. DE-FG02-97ER25308) as well as the Center for Nanoscale Materials, a U.S. Department of Energy, Office of Science, Office of Basic Energy Sciences User Facility (Contract No. DE-AC02-06CH11357).

\section{Appendix: Molecular-Dynamics Simulations} We performed non-equilibrium MD simulations using the LAMMPS code \cite{LAMMPS}. Our system consisted of a Lennard-Jones (LJ) fluid \cite{Allen} of density $\rho_\text{ave}$, atomic mass 16 g/mol, length scale $\sigma=3.15 \AA$, and energy scale $\varepsilon_f=0.15$ kcal/mol, in a plane-Couette setup of channel width $50\text{ }\AA$, with periodic boundary conditions in all directions. We verified that our conclusions are not affected by channel width, by running a small subset of simulations in channels of width $25\text{ }\AA$ and $100\text{ }\AA$. In the remainder of this Appendix, we describe our setup in detail, non-dimensionalizing using the dimensions provided above.
 
The fluid is confined between two rigid sheets of graphene at fixed $z$-coordinate, which move in opposite directions at fixed velocity $u_w$. In each simulation, after an equilibration period of $3\cdot 10^3$ with a Nos\'e-Hoover thermostat \cite{NH_Hoover} at a temperature of 5 (unless otherwise specified), fluid velocities are averaged over a time period of $6\cdot 10^3$ to obtain a velocity profile as a function of $z$-coordinate; to reduce variance in low-signal simulations \cite{jcperror}, the averaging period was extended to $3\cdot 10^4$ if $u_s\leq 0.01$. The use of a thermostat is necessary in these calculations in order to regulate temperature in the presence of viscous heating. For a small subset of simulations, we verified that an alternative choice of the thermostat (in particular a Berendsen thermostat \cite{Berendsen}) does not affect our conclusions. We also verified that the results obtained using no thermostat (simulations in the microcanonical ensemble) agree with the results from both thermostats, provided that the shear rate is low (and so viscous heating is negligible). 

All simulations were performed in the regime $\dot\gamma\tau < 0.07$, where it is well known that simple fluids have shear-rate-independent viscosity \cite{ThompsonTroian, PriezjevSlip}. We determine the slip velocity by fitting a line to the fluid velocity profile away from the solid boundaries (which can induce strong inhomogeneities in the fluid density \cite{FluidsInCNTs, FirstLayerDensity}) and extrapolating to the locations of the walls. In our simulations, we fit the velocity profile over the central region of the channel, at least 3 distance units away from the walls.

\bibliography{Liquid_Slip_at_Solid_Interfaces}

%merlin.mbs apsrev4-1.bst 2010-07-25 4.21a (PWD, AO, DPC) hacked
%Control: key (0)
%Control: author (0) dotless jnrlst
%Control: editor formatted (1) identically to author
%Control: production of article title (0) allowed
%Control: page (1) range
%Control: year (0) verbatim
%Control: production of eprint (0) enabled
\begin{thebibliography}{52}%
\makeatletter
\providecommand \@ifxundefined [1]{%
 \@ifx{#1\undefined}
}%
\providecommand \@ifnum [1]{%
 \ifnum #1\expandafter \@firstoftwo
 \else \expandafter \@secondoftwo
 \fi
}%
\providecommand \@ifx [1]{%
 \ifx #1\expandafter \@firstoftwo
 \else \expandafter \@secondoftwo
 \fi
}%
\providecommand \natexlab [1]{#1}%
\providecommand \enquote  [1]{``#1''}%
\providecommand \bibnamefont  [1]{#1}%
\providecommand \bibfnamefont [1]{#1}%
\providecommand \citenamefont [1]{#1}%
\providecommand \href@noop [0]{\@secondoftwo}%
\providecommand \href [0]{\begingroup \@sanitize@url \@href}%
\providecommand \@href[1]{\@@startlink{#1}\@@href}%
\providecommand \@@href[1]{\endgroup#1\@@endlink}%
\providecommand \@sanitize@url [0]{\catcode `\\12\catcode `\$12\catcode
  `\&12\catcode `\#12\catcode `\^12\catcode `\_12\catcode `\%12\relax}%
\providecommand \@@startlink[1]{}%
\providecommand \@@endlink[0]{}%
\providecommand \url  [0]{\begingroup\@sanitize@url \@url }%
\providecommand \@url [1]{\endgroup\@href {#1}{\urlprefix }}%
\providecommand \urlprefix  [0]{URL }%
\providecommand \Eprint [0]{\href }%
\providecommand \doibase [0]{http://dx.doi.org/}%
\providecommand \selectlanguage [0]{\@gobble}%
\providecommand \bibinfo  [0]{\@secondoftwo}%
\providecommand \bibfield  [0]{\@secondoftwo}%
\providecommand \translation [1]{[#1]}%
\providecommand \BibitemOpen [0]{}%
\providecommand \bibitemStop [0]{}%
\providecommand \bibitemNoStop [0]{.\EOS\space}%
\providecommand \EOS [0]{\spacefactor3000\relax}%
\providecommand \BibitemShut  [1]{\csname bibitem#1\endcsname}%
\let\auto@bib@innerbib\@empty
%</preamble>
\bibitem [{\citenamefont {Hummer}\ \emph {et~al.}(2001)\citenamefont {Hummer},
  \citenamefont {Rasaiah},\ and\ \citenamefont {Noworyta}}]{Hummer}%
  \BibitemOpen
  \bibfield  {author} {\bibinfo {author} {\bibfnamefont {Gerhard}\ \bibnamefont
  {Hummer}}, \bibinfo {author} {\bibfnamefont {J.~C.}\ \bibnamefont {Rasaiah}},
  \ and\ \bibinfo {author} {\bibfnamefont {J.~P.}\ \bibnamefont {Noworyta}},\
  }\bibfield  {title} {\enquote {\bibinfo {title} {Water conduction through the
  hydrophobic channel of a carbon nanotube},}\ }\href {\doibase
  10.1038/35102535} {\bibfield  {journal} {\bibinfo  {journal} {Nature}\
  }\textbf {\bibinfo {volume} {414}},\ \bibinfo {pages} {188 -- 190} (\bibinfo
  {year} {2001})}\BibitemShut {NoStop}%
\bibitem [{\citenamefont {Majumder}\ \emph {et~al.}(2005)\citenamefont
  {Majumder}, \citenamefont {Chopra}, \citenamefont {Andrews},\ and\
  \citenamefont {Hinds}}]{Majumder}%
  \BibitemOpen
  \bibfield  {author} {\bibinfo {author} {\bibfnamefont {Mainak}\ \bibnamefont
  {Majumder}}, \bibinfo {author} {\bibfnamefont {Nitin}\ \bibnamefont
  {Chopra}}, \bibinfo {author} {\bibfnamefont {Rodney}\ \bibnamefont
  {Andrews}}, \ and\ \bibinfo {author} {\bibfnamefont {Bruce~J.}\ \bibnamefont
  {Hinds}},\ }\bibfield  {title} {\enquote {\bibinfo {title} {Nanoscale
  hydrodynamics: Enhanced flow in carbon nanotubes},}\ }\href {\doibase
  10.1038/438044a} {\bibfield  {journal} {\bibinfo  {journal} {Nature}\
  }\textbf {\bibinfo {volume} {438}},\ \bibinfo {pages} {44} (\bibinfo {year}
  {2005})}\BibitemShut {NoStop}%
\bibitem [{\citenamefont {Thomas}\ and\ \citenamefont
  {McGaughey}(2009)}]{ThomasMcGaugheySubcontinuum}%
  \BibitemOpen
  \bibfield  {author} {\bibinfo {author} {\bibfnamefont {John~A.}\ \bibnamefont
  {Thomas}}\ and\ \bibinfo {author} {\bibfnamefont {Alan J.~H.}\ \bibnamefont
  {McGaughey}},\ }\bibfield  {title} {\enquote {\bibinfo {title} {Water flow in
  carbon nanotubes: Transition to subcontinuum transport},}\ }\href {\doibase
  10.1103/PhysRevLett.102.184502} {\bibfield  {journal} {\bibinfo  {journal}
  {Phys. Rev. Lett.}\ }\textbf {\bibinfo {volume} {102}},\ \bibinfo {pages}
  {184502} (\bibinfo {year} {2009})}\BibitemShut {NoStop}%
\bibitem [{\citenamefont {Wang}\ and\ \citenamefont
  {Hadjiconstantinou}(2018)}]{FirstLayerDiffusivity}%
  \BibitemOpen
  \bibfield  {author} {\bibinfo {author} {\bibfnamefont {Gerald~J.}\
  \bibnamefont {Wang}}\ and\ \bibinfo {author} {\bibfnamefont {Nicolas~G.}\
  \bibnamefont {Hadjiconstantinou}},\ }\bibfield  {title} {\enquote {\bibinfo
  {title} {Layered fluid structure and anomalous diffusion under
  nanoconfinement},}\ }\href {\doibase 10.1021/acs.langmuir.8b01540} {\bibfield
   {journal} {\bibinfo  {journal} {Langmuir}\ }\textbf {\bibinfo {volume}
  {34}},\ \bibinfo {pages} {6976--6982} (\bibinfo {year} {2018})},\ \bibinfo
  {note} {pMID: 29775320},\ \Eprint
  {http://arxiv.org/abs/https://doi.org/10.1021/acs.langmuir.8b01540}
  {https://doi.org/10.1021/acs.langmuir.8b01540} \BibitemShut {NoStop}%
\bibitem [{\citenamefont {Alexeev}\ \emph {et~al.}(2015)\citenamefont
  {Alexeev}, \citenamefont {Chen}, \citenamefont {Walther}, \citenamefont
  {Giapis}, \citenamefont {Angelikopoulos},\ and\ \citenamefont
  {Koumoutsakos}}]{FLGKapitza}%
  \BibitemOpen
  \bibfield  {author} {\bibinfo {author} {\bibfnamefont {D.}~\bibnamefont
  {Alexeev}}, \bibinfo {author} {\bibfnamefont {J.}~\bibnamefont {Chen}},
  \bibinfo {author} {\bibfnamefont {J.~H.}\ \bibnamefont {Walther}}, \bibinfo
  {author} {\bibfnamefont {K.~P.}\ \bibnamefont {Giapis}}, \bibinfo {author}
  {\bibfnamefont {P.}~\bibnamefont {Angelikopoulos}}, \ and\ \bibinfo {author}
  {\bibfnamefont {P.}~\bibnamefont {Koumoutsakos}},\ }\bibfield  {title}
  {\enquote {\bibinfo {title} {Kapitza resistance between few-layer graphene
  and water: Liquid layering effects},}\ }\href {\doibase
  10.1021/acs.nanolett.5b03024} {\bibfield  {journal} {\bibinfo  {journal}
  {Nano Lett.}\ }\textbf {\bibinfo {volume} {15}},\ \bibinfo {pages}
  {\hspace{0.5mm} 5744--5749} (\bibinfo {year} {2015})}\BibitemShut {NoStop}%
\bibitem [{\citenamefont {Humplik}\ \emph {et~al.}(2011)\citenamefont
  {Humplik}, \citenamefont {Lee}, \citenamefont {O'Hern}, \citenamefont
  {Fellman}, \citenamefont {Baig}, \citenamefont {Hassan}, \citenamefont
  {Atieh}, \citenamefont {Rahman}, \citenamefont {Laoui}, \citenamefont
  {Karnik},\ and\ \citenamefont {Wang}}]{KarnikDesalReview}%
  \BibitemOpen
  \bibfield  {author} {\bibinfo {author} {\bibfnamefont {T.}~\bibnamefont
  {Humplik}}, \bibinfo {author} {\bibfnamefont {J.}~\bibnamefont {Lee}},
  \bibinfo {author} {\bibfnamefont {S.~C.}\ \bibnamefont {O'Hern}}, \bibinfo
  {author} {\bibfnamefont {B.~A.}\ \bibnamefont {Fellman}}, \bibinfo {author}
  {\bibfnamefont {M.~A.}\ \bibnamefont {Baig}}, \bibinfo {author}
  {\bibfnamefont {S.~F.}\ \bibnamefont {Hassan}}, \bibinfo {author}
  {\bibfnamefont {M.~A.}\ \bibnamefont {Atieh}}, \bibinfo {author}
  {\bibfnamefont {F.}~\bibnamefont {Rahman}}, \bibinfo {author} {\bibfnamefont
  {T.}~\bibnamefont {Laoui}}, \bibinfo {author} {\bibfnamefont
  {R.}~\bibnamefont {Karnik}}, \ and\ \bibinfo {author} {\bibfnamefont {E.~N.}\
  \bibnamefont {Wang}},\ }\bibfield  {title} {\enquote {\bibinfo {title}
  {Nanostructured materials for water desalination},}\ }\href
  {http://stacks.iop.org/0957-4484/22/i=29/a=292001} {\bibfield  {journal}
  {\bibinfo  {journal} {Nanotechnology}\ }\textbf {\bibinfo {volume} {22}},\
  \bibinfo {pages} {292001} (\bibinfo {year} {2011})}\BibitemShut {NoStop}%
\bibitem [{\citenamefont {Wei}\ \emph {et~al.}(2017)\citenamefont {Wei},
  \citenamefont {Zhang},\ and\ \citenamefont
  {Luo}}]{TengfeiLuoEnergyInterfaces}%
  \BibitemOpen
  \bibfield  {author} {\bibinfo {author} {\bibfnamefont {Xingfei}\ \bibnamefont
  {Wei}}, \bibinfo {author} {\bibfnamefont {Teng}\ \bibnamefont {Zhang}}, \
  and\ \bibinfo {author} {\bibfnamefont {Tengfei}\ \bibnamefont {Luo}},\
  }\bibfield  {title} {\enquote {\bibinfo {title} {Thermal energy transport
  across hard-soft interfaces},}\ }\href {\doibase
  10.1021/acsenergylett.7b00570} {\bibfield  {journal} {\bibinfo  {journal}
  {ACS Energy Letters}\ }\textbf {\bibinfo {volume} {2}},\ \bibinfo {pages}
  {2283--2292} (\bibinfo {year} {2017})},\ \Eprint
  {http://arxiv.org/abs/https://doi.org/10.1021/acsenergylett.7b00570}
  {https://doi.org/10.1021/acsenergylett.7b00570} \BibitemShut {NoStop}%
\bibitem [{\citenamefont {Wang}\ and\ \citenamefont
  {Hadjiconstantinou}(2015)}]{FluidsInCNTs}%
  \BibitemOpen
  \bibfield  {author} {\bibinfo {author} {\bibfnamefont {Gerald~J.}\
  \bibnamefont {Wang}}\ and\ \bibinfo {author} {\bibfnamefont {Nicolas~G.}\
  \bibnamefont {Hadjiconstantinou}},\ }\bibfield  {title} {\enquote {\bibinfo
  {title} {Why are fluid densities so low in carbon nanotubes?}}\ }\href@noop
  {} {\bibfield  {journal} {\bibinfo  {journal} {Phys. Fluids}\ }\textbf
  {\bibinfo {volume} {27}},\ \bibinfo {eid} {052006} (\bibinfo {year}
  {2015})}\BibitemShut {NoStop}%
\bibitem [{\citenamefont {Thompson}\ and\ \citenamefont
  {Robbins}(1990)}]{ThompsonRobbinsStickSlip}%
  \BibitemOpen
  \bibfield  {author} {\bibinfo {author} {\bibfnamefont {Peter~A.}\
  \bibnamefont {Thompson}}\ and\ \bibinfo {author} {\bibfnamefont {Mark~O.}\
  \bibnamefont {Robbins}},\ }\bibfield  {title} {\enquote {\bibinfo {title}
  {Origin of stick-slip motion in boundary lubrication},}\ }\href@noop {}
  {\bibfield  {journal} {\bibinfo  {journal} {Science}\ }\textbf {\bibinfo
  {volume} {250}},\ \bibinfo {pages} {792--794} (\bibinfo {year}
  {1990})}\BibitemShut {NoStop}%
\bibitem [{\citenamefont {Chen}\ \emph {et~al.}(2014)\citenamefont {Chen},
  \citenamefont {Walther},\ and\ \citenamefont
  {Koumoutsakos}}]{StrainEngineeringFLG}%
  \BibitemOpen
  \bibfield  {author} {\bibinfo {author} {\bibfnamefont {Jie}\ \bibnamefont
  {Chen}}, \bibinfo {author} {\bibfnamefont {Jens~H.}\ \bibnamefont {Walther}},
  \ and\ \bibinfo {author} {\bibfnamefont {Petros}\ \bibnamefont
  {Koumoutsakos}},\ }\bibfield  {title} {\enquote {\bibinfo {title} {Strain
  engineering of kapitza resistance in few-layer graphene},}\ }\href {\doibase
  10.1021/nl404182k} {\bibfield  {journal} {\bibinfo  {journal} {Nano Lett.}\
  }\textbf {\bibinfo {volume} {14}},\ \bibinfo {pages} {819--825} (\bibinfo
  {year} {2014})}\BibitemShut {NoStop}%
\bibitem [{\citenamefont {Thompson}\ and\ \citenamefont
  {Troian}(1997)}]{ThompsonTroian}%
  \BibitemOpen
  \bibfield  {author} {\bibinfo {author} {\bibfnamefont {Peter~A.}\
  \bibnamefont {Thompson}}\ and\ \bibinfo {author} {\bibfnamefont {Sandra~M.}\
  \bibnamefont {Troian}},\ }\bibfield  {title} {\enquote {\bibinfo {title} {{A
  general boundary condition for liquid flow at solid surfaces}},}\ }\href
  {http://dx.doi.org/10.1038/38686 10.1038/38686} {\bibfield  {journal}
  {\bibinfo  {journal} {Nature}\ }\textbf {\bibinfo {volume} {389}},\ \bibinfo
  {pages} {360} (\bibinfo {year} {1997})}\BibitemShut {NoStop}%
\bibitem [{\citenamefont {Vinogradova}(1995)}]{VinogradovaSlip}%
  \BibitemOpen
  \bibfield  {author} {\bibinfo {author} {\bibfnamefont {Olga~I.}\ \bibnamefont
  {Vinogradova}},\ }\bibfield  {title} {\enquote {\bibinfo {title} {Drainage of
  a thin liquid film confined between hydrophobic surfaces},}\ }\href {\doibase
  10.1021/la00006a059} {\bibfield  {journal} {\bibinfo  {journal} {Langmuir}\
  }\textbf {\bibinfo {volume} {11}},\ \bibinfo {pages} {2213--2220} (\bibinfo
  {year} {1995})},\ \Eprint
  {http://arxiv.org/abs/https://doi.org/10.1021/la00006a059}
  {https://doi.org/10.1021/la00006a059} \BibitemShut {NoStop}%
\bibitem [{\citenamefont {de~Gennes}(2002)}]{deGennesSlip}%
  \BibitemOpen
  \bibfield  {author} {\bibinfo {author} {\bibfnamefont {P.~G.}\ \bibnamefont
  {de~Gennes}},\ }\bibfield  {title} {\enquote {\bibinfo {title} {On fluid/wall
  slippage},}\ }\href {\doibase 10.1021/la0116342} {\bibfield  {journal}
  {\bibinfo  {journal} {Langmuir}\ }\textbf {\bibinfo {volume} {18}},\ \bibinfo
  {pages} {3413--3414} (\bibinfo {year} {2002})},\ \Eprint
  {http://arxiv.org/abs/https://doi.org/10.1021/la0116342}
  {https://doi.org/10.1021/la0116342} \BibitemShut {NoStop}%
\bibitem [{\citenamefont {Barrat}\ and\ \citenamefont
  {Bocquet}(1999{\natexlab{a}})}]{FirstLayerSlip}%
  \BibitemOpen
  \bibfield  {author} {\bibinfo {author} {\bibfnamefont {J.-L.}\ \bibnamefont
  {Barrat}}\ and\ \bibinfo {author} {\bibfnamefont {L.}~\bibnamefont
  {Bocquet}},\ }\bibfield  {title} {\enquote {\bibinfo {title} {Influence of
  wetting properties on hydrodynamic boundary conditions at a fluid/solid
  interface},}\ }\href@noop {} {\bibfield  {journal} {\bibinfo  {journal}
  {Faraday Discuss.}\ }\textbf {\bibinfo {volume} {112}},\ \bibinfo {pages}
  {119--127} (\bibinfo {year} {1999}{\natexlab{a}})}\BibitemShut {NoStop}%
\bibitem [{\citenamefont {Lichter}\ \emph {et~al.}(2004)\citenamefont
  {Lichter}, \citenamefont {Roxin},\ and\ \citenamefont
  {Mandre}}]{Lichter_vdFK}%
  \BibitemOpen
  \bibfield  {author} {\bibinfo {author} {\bibfnamefont {Seth}\ \bibnamefont
  {Lichter}}, \bibinfo {author} {\bibfnamefont {Alex}\ \bibnamefont {Roxin}}, \
  and\ \bibinfo {author} {\bibfnamefont {Shreyas}\ \bibnamefont {Mandre}},\
  }\bibfield  {title} {\enquote {\bibinfo {title} {Mechanisms for liquid slip
  at solid surfaces},}\ }\href {\doibase 10.1103/PhysRevLett.93.086001}
  {\bibfield  {journal} {\bibinfo  {journal} {Phys. Rev. Lett.}\ }\textbf
  {\bibinfo {volume} {93}},\ \bibinfo {pages} {086001} (\bibinfo {year}
  {2004})}\BibitemShut {NoStop}%
\bibitem [{\citenamefont {Priezjev}\ and\ \citenamefont
  {Troian}(2006)}]{PriezjevTroian}%
  \BibitemOpen
  \bibfield  {author} {\bibinfo {author} {\bibfnamefont {Nikolai~V.}\
  \bibnamefont {Priezjev}}\ and\ \bibinfo {author} {\bibfnamefont {Sandra~M.}\
  \bibnamefont {Troian}},\ }\bibfield  {title} {\enquote {\bibinfo {title}
  {Influence of periodic wall roughness on the slip behaviour at liquid/solid
  interfaces: molecular-scale simulations versus continuum predictions},}\
  }\href {\doibase 10.1017/S0022112006009086} {\bibfield  {journal} {\bibinfo
  {journal} {Journal of Fluid Mechanics}\ }\textbf {\bibinfo {volume} {554}},\
  \bibinfo {pages} {25--46} (\bibinfo {year} {2006})}\BibitemShut {NoStop}%
\bibitem [{\citenamefont {Shu}\ \emph {et~al.}(2017)\citenamefont {Shu},
  \citenamefont {{Bin Melvin Teo}},\ and\ \citenamefont {{Kong
  Chan}}}]{ShuTeoChanReview}%
  \BibitemOpen
  \bibfield  {author} {\bibinfo {author} {\bibfnamefont {Jian-Jun}\
  \bibnamefont {Shu}}, \bibinfo {author} {\bibfnamefont {Ji}~\bibnamefont {{Bin
  Melvin Teo}}}, \ and\ \bibinfo {author} {\bibfnamefont {Weng}\ \bibnamefont
  {{Kong Chan}}},\ }\bibfield  {title} {\enquote {\bibinfo {title} {{Fluid
  Velocity Slip and Temperature Jump at a Solid Surface}},}\ }\href
  {http://dx.doi.org/10.1115/1.4036191} {\bibfield  {journal} {\bibinfo
  {journal} {Applied Mechanics Reviews}\ }\textbf {\bibinfo {volume} {69}},\
  \bibinfo {pages} {20801--20813} (\bibinfo {year} {2017})}\BibitemShut
  {NoStop}%
\bibitem [{\citenamefont
  {Priezjev}(2007{\natexlab{a}})}]{RateDependentSlipBoundary}%
  \BibitemOpen
  \bibfield  {author} {\bibinfo {author} {\bibfnamefont {Nikolai~V.}\
  \bibnamefont {Priezjev}},\ }\bibfield  {title} {\enquote {\bibinfo {title}
  {Rate-dependent slip boundary conditions for simple fluids},}\ }\href
  {\doibase 10.1103/PhysRevE.75.051605} {\bibfield  {journal} {\bibinfo
  {journal} {Phys. Rev. E}\ }\textbf {\bibinfo {volume} {75}},\ \bibinfo
  {pages} {051605} (\bibinfo {year} {2007}{\natexlab{a}})}\BibitemShut
  {NoStop}%
\bibitem [{\citenamefont {Sone}(2007)}]{Sone2007}%
  \BibitemOpen
  \bibfield  {author} {\bibinfo {author} {\bibfnamefont {Y.}~\bibnamefont
  {Sone}},\ }\href@noop {} {\emph {\bibinfo {title} {Molecular Gas Dynamics:
  Theory, Techniques, and Applications}}}\ (\bibinfo  {publisher}
  {Birkh\"auser},\ \bibinfo {year} {2007})\BibitemShut {NoStop}%
\bibitem [{\citenamefont {Hadjiconstantinou}(2006)}]{pof2006}%
  \BibitemOpen
  \bibfield  {author} {\bibinfo {author} {\bibfnamefont {Nicolas~G.}\
  \bibnamefont {Hadjiconstantinou}},\ }\bibfield  {title} {\enquote {\bibinfo
  {title} {The limits of {N}avier-{S}tokes theory and kinetic extensions for
  describing small-scale gaseous hydrodynamics},}\ }\href {\doibase
  10.1063/1.2393436} {\bibfield  {journal} {\bibinfo  {journal} {Physics of
  Fluids}\ }\textbf {\bibinfo {volume} {18}},\ \bibinfo {pages} {111301}
  (\bibinfo {year} {2006})},\ \Eprint
  {http://arxiv.org/abs/https://doi.org/10.1063/1.2393436}
  {https://doi.org/10.1063/1.2393436} \BibitemShut {NoStop}%
\bibitem [{\citenamefont {P\'eraud}\ and\ \citenamefont
  {Hadjiconstantinou}(2016)}]{JP2016}%
  \BibitemOpen
  \bibfield  {author} {\bibinfo {author} {\bibfnamefont {Jean-Philippe~M.}\
  \bibnamefont {P\'eraud}}\ and\ \bibinfo {author} {\bibfnamefont {Nicolas~G.}\
  \bibnamefont {Hadjiconstantinou}},\ }\bibfield  {title} {\enquote {\bibinfo
  {title} {Extending the range of validity of fourier's law into the kinetic
  transport regime via asymptotic solution of the phonon boltzmann transport
  equation},}\ }\href {\doibase 10.1103/PhysRevB.93.045424} {\bibfield
  {journal} {\bibinfo  {journal} {Phys. Rev. B}\ }\textbf {\bibinfo {volume}
  {93}},\ \bibinfo {pages} {045424} (\bibinfo {year} {2016})}\BibitemShut
  {NoStop}%
\bibitem [{\citenamefont {Navier}(1823)}]{NavierSlip}%
  \BibitemOpen
  \bibfield  {author} {\bibinfo {author} {\bibfnamefont {Claude}\ \bibnamefont
  {Navier}},\ }\bibfield  {title} {\enquote {\bibinfo {title} {M\'emoire sur
  les lois du mouvement des fluides},}\ }\href@noop {} {\bibfield  {journal}
  {\bibinfo  {journal} {M\'emoires de l'Acad\'emie Royale des Sciences de
  l'Institut de France}\ }\textbf {\bibinfo {volume} {VI}},\ \bibinfo {pages}
  {389 -- 440} (\bibinfo {year} {1823})}\BibitemShut {NoStop}%
\bibitem [{\citenamefont {Bocquet}\ and\ \citenamefont
  {Barrat}(2013)}]{BocquetBarratGKFriction}%
  \BibitemOpen
  \bibfield  {author} {\bibinfo {author} {\bibfnamefont {Lyd\'eric}\
  \bibnamefont {Bocquet}}\ and\ \bibinfo {author} {\bibfnamefont {Jean-Louis}\
  \bibnamefont {Barrat}},\ }\bibfield  {title} {\enquote {\bibinfo {title} {On
  the green-kubo relationship for the liquid-solid friction coefficient},}\
  }\href {\doibase 10.1063/1.4816006} {\bibfield  {journal} {\bibinfo
  {journal} {The Journal of Chemical Physics}\ }\textbf {\bibinfo {volume}
  {139}},\ \bibinfo {pages} {044704} (\bibinfo {year} {2013})},\ \Eprint
  {http://arxiv.org/abs/https://doi.org/10.1063/1.4816006}
  {https://doi.org/10.1063/1.4816006} \BibitemShut {NoStop}%
\bibitem [{\citenamefont {Blake}\ and\ \citenamefont
  {Haynes}(1969)}]{BlakeHaynes}%
  \BibitemOpen
  \bibfield  {author} {\bibinfo {author} {\bibfnamefont {T.D}\ \bibnamefont
  {Blake}}\ and\ \bibinfo {author} {\bibfnamefont {J.M}\ \bibnamefont
  {Haynes}},\ }\bibfield  {title} {\enquote {\bibinfo {title} {Kinetics of
  liquid/liquid displacement},}\ }\href {\doibase
  https://doi.org/10.1016/0021-9797(69)90411-1} {\bibfield  {journal} {\bibinfo
   {journal} {Journal of Colloid and Interface Science}\ }\textbf {\bibinfo
  {volume} {30}},\ \bibinfo {pages} {421 -- 423} (\bibinfo {year}
  {1969})}\BibitemShut {NoStop}%
\bibitem [{\citenamefont {Lichter}\ \emph {et~al.}(2007)\citenamefont
  {Lichter}, \citenamefont {Martini}, \citenamefont {Snurr},\ and\
  \citenamefont {Wang}}]{LichterSlipRateProcess}%
  \BibitemOpen
  \bibfield  {author} {\bibinfo {author} {\bibfnamefont {Seth}\ \bibnamefont
  {Lichter}}, \bibinfo {author} {\bibfnamefont {Ashlie}\ \bibnamefont
  {Martini}}, \bibinfo {author} {\bibfnamefont {Randall~Q.}\ \bibnamefont
  {Snurr}}, \ and\ \bibinfo {author} {\bibfnamefont {Qian}\ \bibnamefont
  {Wang}},\ }\bibfield  {title} {\enquote {\bibinfo {title} {Liquid slip in
  nanoscale channels as a rate process},}\ }\href {\doibase
  10.1103/PhysRevLett.98.226001} {\bibfield  {journal} {\bibinfo  {journal}
  {Phys. Rev. Lett.}\ }\textbf {\bibinfo {volume} {98}},\ \bibinfo {pages}
  {226001} (\bibinfo {year} {2007})}\BibitemShut {NoStop}%
\bibitem [{\citenamefont {Martini}\ \emph
  {et~al.}(2008{\natexlab{a}})\citenamefont {Martini}, \citenamefont {Roxin},
  \citenamefont {Snurr}, \citenamefont {Wang},\ and\ \citenamefont
  {Lichter}}]{MartiniLichterJFM}%
  \BibitemOpen
  \bibfield  {author} {\bibinfo {author} {\bibfnamefont {A.}~\bibnamefont
  {Martini}}, \bibinfo {author} {\bibfnamefont {A.}~\bibnamefont {Roxin}},
  \bibinfo {author} {\bibfnamefont {R.~Q.}\ \bibnamefont {Snurr}}, \bibinfo
  {author} {\bibfnamefont {Q.}~\bibnamefont {Wang}}, \ and\ \bibinfo {author}
  {\bibfnamefont {S.}~\bibnamefont {Lichter}},\ }\bibfield  {title} {\enquote
  {\bibinfo {title} {Molecular mechanisms of liquid slip},}\ }\href {\doibase
  10.1017/S0022112008000475} {\bibfield  {journal} {\bibinfo  {journal}
  {Journal of Fluid Mechanics}\ }\textbf {\bibinfo {volume} {600}},\ \bibinfo
  {pages} {257–269} (\bibinfo {year} {2008}{\natexlab{a}})}\BibitemShut
  {NoStop}%
\bibitem [{\citenamefont {Eyring}(1936)}]{EyringReactionRates}%
  \BibitemOpen
  \bibfield  {author} {\bibinfo {author} {\bibfnamefont {Henry}\ \bibnamefont
  {Eyring}},\ }\bibfield  {title} {\enquote {\bibinfo {title} {Viscosity,
  plasticity, and diffusion as examples of absolute reaction rates},}\ }\href
  {\doibase 10.1063/1.1749836} {\bibfield  {journal} {\bibinfo  {journal} {The
  Journal of Chemical Physics}\ }\textbf {\bibinfo {volume} {4}},\ \bibinfo
  {pages} {283--291} (\bibinfo {year} {1936})},\ \Eprint
  {http://arxiv.org/abs/https://doi.org/10.1063/1.1749836}
  {https://doi.org/10.1063/1.1749836} \BibitemShut {NoStop}%
\bibitem [{\citenamefont {Hirschfelder}\ \emph {et~al.}(1955)\citenamefont
  {Hirschfelder}, \citenamefont {Curtiss},\ and\ \citenamefont
  {Bird}}]{Hirschfelder}%
  \BibitemOpen
  \bibfield  {author} {\bibinfo {author} {\bibfnamefont {J.~O.}\ \bibnamefont
  {Hirschfelder}}, \bibinfo {author} {\bibfnamefont {C.~F.}\ \bibnamefont
  {Curtiss}}, \ and\ \bibinfo {author} {\bibfnamefont {R.~B.}\ \bibnamefont
  {Bird}},\ }\href@noop {} {\emph {\bibinfo {title} {Molecular theory of gases
  and liquids}}}\ (\bibinfo  {publisher} {Wiley},\ \bibinfo {year}
  {1955})\BibitemShut {NoStop}%
\bibitem [{\citenamefont {H\"anggi}\ \emph {et~al.}(1990)\citenamefont
  {H\"anggi}, \citenamefont {Talkner},\ and\ \citenamefont
  {Borkovec}}]{HanggiKramers}%
  \BibitemOpen
  \bibfield  {author} {\bibinfo {author} {\bibfnamefont {Peter}\ \bibnamefont
  {H\"anggi}}, \bibinfo {author} {\bibfnamefont {Peter}\ \bibnamefont
  {Talkner}}, \ and\ \bibinfo {author} {\bibfnamefont {Michal}\ \bibnamefont
  {Borkovec}},\ }\bibfield  {title} {\enquote {\bibinfo {title} {Reaction-rate
  theory: fifty years after kramers},}\ }\href@noop {} {\bibfield  {journal}
  {\bibinfo  {journal} {Rev. Mod. Phys.}\ }\textbf {\bibinfo {volume} {62}},\
  \bibinfo {pages} {251--341} (\bibinfo {year} {1990})}\BibitemShut {NoStop}%
\bibitem [{\citenamefont {Bonn}\ \emph {et~al.}(2009)\citenamefont {Bonn},
  \citenamefont {Eggers}, \citenamefont {Indekeu}, \citenamefont {Meunier},\
  and\ \citenamefont {Rolley}}]{Eggers}%
  \BibitemOpen
  \bibfield  {author} {\bibinfo {author} {\bibfnamefont {Daniel}\ \bibnamefont
  {Bonn}}, \bibinfo {author} {\bibfnamefont {Jens}\ \bibnamefont {Eggers}},
  \bibinfo {author} {\bibfnamefont {Joseph}\ \bibnamefont {Indekeu}}, \bibinfo
  {author} {\bibfnamefont {Jacques}\ \bibnamefont {Meunier}}, \ and\ \bibinfo
  {author} {\bibfnamefont {Etienne}\ \bibnamefont {Rolley}},\ }\bibfield
  {title} {\enquote {\bibinfo {title} {Wetting and spreading},}\ }\href
  {\doibase 10.1103/RevModPhys.81.739} {\bibfield  {journal} {\bibinfo
  {journal} {Rev. Mod. Phys.}\ }\textbf {\bibinfo {volume} {81}},\ \bibinfo
  {pages} {739--805} (\bibinfo {year} {2009})}\BibitemShut {NoStop}%
\bibitem [{\citenamefont {Blake}(2006)}]{BlakeReview}%
  \BibitemOpen
  \bibfield  {author} {\bibinfo {author} {\bibfnamefont {Terence~D.}\
  \bibnamefont {Blake}},\ }\bibfield  {title} {\enquote {\bibinfo {title} {The
  physics of moving wetting lines},}\ }\href {\doibase
  https://doi.org/10.1016/j.jcis.2006.03.051} {\bibfield  {journal} {\bibinfo
  {journal} {Journal of Colloid and Interface Science}\ }\textbf {\bibinfo
  {volume} {299}},\ \bibinfo {pages} {1 -- 13} (\bibinfo {year}
  {2006})}\BibitemShut {NoStop}%
\bibitem [{\citenamefont {Ren}\ and\ \citenamefont {E}(2007)}]{Ren2007}%
  \BibitemOpen
  \bibfield  {author} {\bibinfo {author} {\bibfnamefont {Weiqing}\ \bibnamefont
  {Ren}}\ and\ \bibinfo {author} {\bibfnamefont {Weinan}\ \bibnamefont {E}},\
  }\bibfield  {title} {\enquote {\bibinfo {title} {Boundary conditions for the
  moving contact line problem},}\ }\href {\doibase 10.1063/1.2646754}
  {\bibfield  {journal} {\bibinfo  {journal} {Physics of Fluids}\ }\textbf
  {\bibinfo {volume} {19}},\ \bibinfo {pages} {022101} (\bibinfo {year}
  {2007})},\ \Eprint {http://arxiv.org/abs/https://doi.org/10.1063/1.2646754}
  {https://doi.org/10.1063/1.2646754} \BibitemShut {NoStop}%
\bibitem [{\citenamefont {Wang}\ \emph {et~al.}(2019)\citenamefont {Wang},
  \citenamefont {Damone}, \citenamefont {Benfenati}, \citenamefont {Poesio},
  \citenamefont {Beretta},\ and\ \citenamefont
  {Hadjiconstantinou}}]{ItalianPaper}%
  \BibitemOpen
  \bibfield  {author} {\bibinfo {author} {\bibfnamefont {Gerald~J.}\
  \bibnamefont {Wang}}, \bibinfo {author} {\bibfnamefont {Angelo}\ \bibnamefont
  {Damone}}, \bibinfo {author} {\bibfnamefont {Francesco}\ \bibnamefont
  {Benfenati}}, \bibinfo {author} {\bibfnamefont {Pietro}\ \bibnamefont
  {Poesio}}, \bibinfo {author} {\bibfnamefont {Gian~Paolo}\ \bibnamefont
  {Beretta}}, \ and\ \bibinfo {author} {\bibfnamefont {Nicolas~G.}\
  \bibnamefont {Hadjiconstantinou}},\ }\bibfield  {title} {\enquote {\bibinfo
  {title} {Physics of nanoscale immiscible fluid displacement},}\ }\href@noop
  {} {\bibfield  {journal} {\bibinfo  {journal} {under review}\ } (\bibinfo
  {year} {2019})}\BibitemShut {NoStop}%
\bibitem [{\citenamefont {Brochard-Wyart}\ and\ \citenamefont
  {de~Gennes}(1992)}]{Wyart_and_deGennes}%
  \BibitemOpen
  \bibfield  {author} {\bibinfo {author} {\bibfnamefont {F.}~\bibnamefont
  {Brochard-Wyart}}\ and\ \bibinfo {author} {\bibfnamefont {P.G.}\ \bibnamefont
  {de~Gennes}},\ }\bibfield  {title} {\enquote {\bibinfo {title} {Dynamics of
  partial wetting},}\ }\href {\doibase
  https://doi.org/10.1016/0001-8686(92)80052-Y} {\bibfield  {journal} {\bibinfo
   {journal} {Advances in Colloid and Interface Science}\ }\textbf {\bibinfo
  {volume} {39}},\ \bibinfo {pages} {1 -- 11} (\bibinfo {year}
  {1992})}\BibitemShut {NoStop}%
\bibitem [{\citenamefont {Wang}\ and\ \citenamefont
  {Hadjiconstantinou}(2017)}]{FirstLayerDensity}%
  \BibitemOpen
  \bibfield  {author} {\bibinfo {author} {\bibfnamefont {Gerald~J.}\
  \bibnamefont {Wang}}\ and\ \bibinfo {author} {\bibfnamefont {Nicolas~G.}\
  \bibnamefont {Hadjiconstantinou}},\ }\bibfield  {title} {\enquote {\bibinfo
  {title} {Molecular mechanics and structure of the fluid-solid interface in
  simple fluids},}\ }\href@noop {} {\bibfield  {journal} {\bibinfo  {journal}
  {Phys. Rev. Fluids}\ }\textbf {\bibinfo {volume} {2}},\ \bibinfo {pages}
  {094201} (\bibinfo {year} {2017})}\BibitemShut {NoStop}%
\bibitem [{\citenamefont {Allen}\ and\ \citenamefont
  {Tildesley}(1989)}]{Allen}%
  \BibitemOpen
  \bibfield  {author} {\bibinfo {author} {\bibfnamefont {Michael~P.}\
  \bibnamefont {Allen}}\ and\ \bibinfo {author} {\bibfnamefont {Dominic~J.}\
  \bibnamefont {Tildesley}},\ }\href@noop {} {\emph {\bibinfo {title} {Computer
  Simulation of Liquids}}}\ (\bibinfo  {publisher} {Oxford University Press},\
  \bibinfo {year} {1989})\BibitemShut {NoStop}%
\bibitem [{\citenamefont {Martini}\ \emph
  {et~al.}(2008{\natexlab{b}})\citenamefont {Martini}, \citenamefont {Hsu},
  \citenamefont {Patankar},\ and\ \citenamefont {Lichter}}]{MartiniSlip}%
  \BibitemOpen
  \bibfield  {author} {\bibinfo {author} {\bibfnamefont {Ashlie}\ \bibnamefont
  {Martini}}, \bibinfo {author} {\bibfnamefont {Hua-Yi}\ \bibnamefont {Hsu}},
  \bibinfo {author} {\bibfnamefont {Neelesh~A.}\ \bibnamefont {Patankar}}, \
  and\ \bibinfo {author} {\bibfnamefont {Seth}\ \bibnamefont {Lichter}},\
  }\bibfield  {title} {\enquote {\bibinfo {title} {Slip at high shear rates},}\
  }\href {\doibase 10.1103/PhysRevLett.100.206001} {\bibfield  {journal}
  {\bibinfo  {journal} {Phys. Rev. Lett.}\ }\textbf {\bibinfo {volume} {100}},\
  \bibinfo {pages} {206001} (\bibinfo {year} {2008}{\natexlab{b}})}\BibitemShut
  {NoStop}%
\bibitem [{\citenamefont {Craig}\ \emph {et~al.}(2001)\citenamefont {Craig},
  \citenamefont {Neto},\ and\ \citenamefont {Williams}}]{CraigNetoWilliams}%
  \BibitemOpen
  \bibfield  {author} {\bibinfo {author} {\bibfnamefont {Vincent S.~J.}\
  \bibnamefont {Craig}}, \bibinfo {author} {\bibfnamefont {Chiara}\
  \bibnamefont {Neto}}, \ and\ \bibinfo {author} {\bibfnamefont {David R.~M.}\
  \bibnamefont {Williams}},\ }\bibfield  {title} {\enquote {\bibinfo {title}
  {Shear-dependent boundary slip in an aqueous newtonian liquid},}\ }\href
  {\doibase 10.1103/PhysRevLett.87.054504} {\bibfield  {journal} {\bibinfo
  {journal} {Phys. Rev. Lett.}\ }\textbf {\bibinfo {volume} {87}},\ \bibinfo
  {pages} {054504} (\bibinfo {year} {2001})}\BibitemShut {NoStop}%
\bibitem [{\citenamefont {Zhu}\ and\ \citenamefont
  {Granick}(2001)}]{ZhuGranickSlip}%
  \BibitemOpen
  \bibfield  {author} {\bibinfo {author} {\bibfnamefont {Yingxi}\ \bibnamefont
  {Zhu}}\ and\ \bibinfo {author} {\bibfnamefont {Steve}\ \bibnamefont
  {Granick}},\ }\bibfield  {title} {\enquote {\bibinfo {title} {Rate-dependent
  slip of newtonian liquid at smooth surfaces},}\ }\href {\doibase
  10.1103/PhysRevLett.87.096105} {\bibfield  {journal} {\bibinfo  {journal}
  {Phys. Rev. Lett.}\ }\textbf {\bibinfo {volume} {87}},\ \bibinfo {pages}
  {096105} (\bibinfo {year} {2001})}\BibitemShut {NoStop}%
\bibitem [{\citenamefont {L\'eger}(2003)}]{LegerSlip}%
  \BibitemOpen
  \bibfield  {author} {\bibinfo {author} {\bibfnamefont {Liliane}\ \bibnamefont
  {L\'eger}},\ }\bibfield  {title} {\enquote {\bibinfo {title} {Friction
  mechanisms and interfacial slip at fluid-solid interfaces},}\ }\href
  {http://stacks.iop.org/0953-8984/15/i=1/a=303} {\bibfield  {journal}
  {\bibinfo  {journal} {Journal of Physics: Condensed Matter}\ }\textbf
  {\bibinfo {volume} {15}},\ \bibinfo {pages} {S19} (\bibinfo {year}
  {2003})}\BibitemShut {NoStop}%
\bibitem [{\citenamefont {Khordad}(2008)}]{LJ_Viscosity}%
  \BibitemOpen
  \bibfield  {author} {\bibinfo {author} {\bibfnamefont {R.}~\bibnamefont
  {Khordad}},\ }\bibfield  {title} {\enquote {\bibinfo {title} {Viscosity of
  lennard-jones fluid: Integral equation method},}\ }\href {\doibase
  https://doi.org/10.1016/j.physa.2008.03.025} {\bibfield  {journal} {\bibinfo
  {journal} {Physica A: Statistical Mechanics and its Applications}\ }\textbf
  {\bibinfo {volume} {387}},\ \bibinfo {pages} {4519 -- 4530} (\bibinfo {year}
  {2008})}\BibitemShut {NoStop}%
\bibitem [{\citenamefont {Meier}\ \emph {et~al.}(2004)\citenamefont {Meier},
  \citenamefont {Laesecke},\ and\ \citenamefont {Kabelac}}]{LJTransportCoeff}%
  \BibitemOpen
  \bibfield  {author} {\bibinfo {author} {\bibfnamefont {Karsten}\ \bibnamefont
  {Meier}}, \bibinfo {author} {\bibfnamefont {Arno}\ \bibnamefont {Laesecke}},
  \ and\ \bibinfo {author} {\bibfnamefont {Stephan}\ \bibnamefont {Kabelac}},\
  }\bibfield  {title} {\enquote {\bibinfo {title} {Transport coefficients of
  the lennard-jones model fluid. i. viscosity},}\ }\href {\doibase
  10.1063/1.1770695} {\bibfield  {journal} {\bibinfo  {journal} {The Journal of
  Chemical Physics}\ }\textbf {\bibinfo {volume} {121}},\ \bibinfo {pages}
  {3671--3687} (\bibinfo {year} {2004})},\ \Eprint
  {http://arxiv.org/abs/https://doi.org/10.1063/1.1770695}
  {https://doi.org/10.1063/1.1770695} \BibitemShut {NoStop}%
\bibitem [{\citenamefont {Barrat}\ and\ \citenamefont
  {Bocquet}(1999{\natexlab{b}})}]{BarratBocquet_LargeSlipEffect}%
  \BibitemOpen
  \bibfield  {author} {\bibinfo {author} {\bibfnamefont {Jean-Louis}\
  \bibnamefont {Barrat}}\ and\ \bibinfo {author} {\bibfnamefont {Lyd\'eric}\
  \bibnamefont {Bocquet}},\ }\bibfield  {title} {\enquote {\bibinfo {title}
  {Large slip effect at a nonwetting fluid-solid interface},}\ }\href {\doibase
  10.1103/PhysRevLett.82.4671} {\bibfield  {journal} {\bibinfo  {journal}
  {Phys. Rev. Lett.}\ }\textbf {\bibinfo {volume} {82}},\ \bibinfo {pages}
  {4671--4674} (\bibinfo {year} {1999}{\natexlab{b}})}\BibitemShut {NoStop}%
\bibitem [{\citenamefont {Baudry}\ \emph {et~al.}(2001)\citenamefont {Baudry},
  \citenamefont {Charlaix}, \citenamefont {Tonck},\ and\ \citenamefont
  {Mazuyer}}]{Baudry_ExperimentalSlip}%
  \BibitemOpen
  \bibfield  {author} {\bibinfo {author} {\bibfnamefont {J.}~\bibnamefont
  {Baudry}}, \bibinfo {author} {\bibfnamefont {E.}~\bibnamefont {Charlaix}},
  \bibinfo {author} {\bibfnamefont {A.}~\bibnamefont {Tonck}}, \ and\ \bibinfo
  {author} {\bibfnamefont {D.}~\bibnamefont {Mazuyer}},\ }\bibfield  {title}
  {\enquote {\bibinfo {title} {Experimental evidence for a large slip effect at
  a nonwetting fluid-Âsolid interface},}\ }\href {\doibase
  10.1021/la0009994} {\bibfield  {journal} {\bibinfo  {journal} {Langmuir}\
  }\textbf {\bibinfo {volume} {17}},\ \bibinfo {pages} {5232--5236} (\bibinfo
  {year} {2001})},\ \Eprint
  {http://arxiv.org/abs/https://doi.org/10.1021/la0009994}
  {https://doi.org/10.1021/la0009994} \BibitemShut {NoStop}%
\bibitem [{\citenamefont {Cottin-Bizonne}\ \emph {et~al.}(2002)\citenamefont
  {Cottin-Bizonne}, \citenamefont {Jurine}, \citenamefont {Baudry},
  \citenamefont {Crassous}, \citenamefont {Restagno},\ and\ \citenamefont
  {Charlaix}}]{Cottin-Bizonne_ExperimentalSlip}%
  \BibitemOpen
  \bibfield  {author} {\bibinfo {author} {\bibfnamefont {C.}~\bibnamefont
  {Cottin-Bizonne}}, \bibinfo {author} {\bibfnamefont {S.}~\bibnamefont
  {Jurine}}, \bibinfo {author} {\bibfnamefont {J.}~\bibnamefont {Baudry}},
  \bibinfo {author} {\bibfnamefont {J.}~\bibnamefont {Crassous}}, \bibinfo
  {author} {\bibfnamefont {F.}~\bibnamefont {Restagno}}, \ and\ \bibinfo
  {author} {\bibfnamefont {{\'E}.}~\bibnamefont {Charlaix}},\ }\bibfield
  {title} {\enquote {\bibinfo {title} {Nanorheology: An investigation of the
  boundary condition at hydrophobic and hydrophilic interfaces},}\ }\href
  {\doibase 10.1140/epje/i2001-10112-9} {\bibfield  {journal} {\bibinfo
  {journal} {The European Physical Journal E}\ }\textbf {\bibinfo {volume}
  {9}},\ \bibinfo {pages} {47--53} (\bibinfo {year} {2002})}\BibitemShut
  {NoStop}%
\bibitem [{\citenamefont {Priezjev}\ \emph {et~al.}(2005)\citenamefont
  {Priezjev}, \citenamefont {Darhuber},\ and\ \citenamefont
  {Troian}}]{PriezjevTroianPatternedWettability}%
  \BibitemOpen
  \bibfield  {author} {\bibinfo {author} {\bibfnamefont {Nikolai~V.}\
  \bibnamefont {Priezjev}}, \bibinfo {author} {\bibfnamefont {Anton~A.}\
  \bibnamefont {Darhuber}}, \ and\ \bibinfo {author} {\bibfnamefont
  {Sandra~M.}\ \bibnamefont {Troian}},\ }\bibfield  {title} {\enquote {\bibinfo
  {title} {Slip behavior in liquid films on surfaces of patterned wettability:
  Comparison between continuum and molecular dynamics simulations},}\ }\href
  {\doibase 10.1103/PhysRevE.71.041608} {\bibfield  {journal} {\bibinfo
  {journal} {Phys. Rev. E}\ }\textbf {\bibinfo {volume} {71}},\ \bibinfo
  {pages} {041608} (\bibinfo {year} {2005})}\BibitemShut {NoStop}%
\bibitem [{\citenamefont {Priezjev}(2007{\natexlab{b}})}]{PriezjevSlip}%
  \BibitemOpen
  \bibfield  {author} {\bibinfo {author} {\bibfnamefont {Nikolai~V.}\
  \bibnamefont {Priezjev}},\ }\bibfield  {title} {\enquote {\bibinfo {title}
  {Effect of surface roughness on rate-dependent slip in simple fluids},}\
  }\href {\doibase 10.1063/1.2796172} {\bibfield  {journal} {\bibinfo
  {journal} {The Journal of Chemical Physics}\ }\textbf {\bibinfo {volume}
  {127}},\ \bibinfo {pages} {144708} (\bibinfo {year} {2007}{\natexlab{b}})},\
  \Eprint {http://arxiv.org/abs/https://doi.org/10.1063/1.2796172}
  {https://doi.org/10.1063/1.2796172} \BibitemShut {NoStop}%
\bibitem [{\citenamefont {Heyes}(1985)}]{HeyesShearThinning}%
  \BibitemOpen
  \bibfield  {author} {\bibinfo {author} {\bibfnamefont {David~M.}\
  \bibnamefont {Heyes}},\ }\bibfield  {title} {\enquote {\bibinfo {title}
  {Shear thinning of the {L}ennard-{J}ones fluid by molecular dynamics},}\
  }\href {\doibase https://doi.org/10.1016/0378-4371(85)90144-X} {\bibfield
  {journal} {\bibinfo  {journal} {Physica A: Statistical Mechanics and its
  Applications}\ }\textbf {\bibinfo {volume} {133}},\ \bibinfo {pages} {473 --
  496} (\bibinfo {year} {1985})}\BibitemShut {NoStop}%
\bibitem [{\citenamefont {Plimpton}(1995)}]{LAMMPS}%
  \BibitemOpen
  \bibfield  {author} {\bibinfo {author} {\bibfnamefont {Steve}\ \bibnamefont
  {Plimpton}},\ }\bibfield  {title} {\enquote {\bibinfo {title} {Fast parallel
  algorithms for short-range molecular dynamics},}\ }\href {\doibase
  http://lammps.sandia.gov} {\bibfield  {journal} {\bibinfo  {journal} {J.
  Comp. Phys.}\ }\textbf {\bibinfo {volume} {117}},\ \bibinfo {pages} {1 -- 19}
  (\bibinfo {year} {1995})}\BibitemShut {NoStop}%
\bibitem [{\citenamefont {Hoover}(1985)}]{NH_Hoover}%
  \BibitemOpen
  \bibfield  {author} {\bibinfo {author} {\bibfnamefont {William~G.}\
  \bibnamefont {Hoover}},\ }\bibfield  {title} {\enquote {\bibinfo {title}
  {Canonical dynamics: Equilibrium phase-space distributions},}\ }\href@noop {}
  {\bibfield  {journal} {\bibinfo  {journal} {Phys. Rev. A}\ }\textbf {\bibinfo
  {volume} {31}},\ \bibinfo {pages} {1695--1697} (\bibinfo {year}
  {1985})}\BibitemShut {NoStop}%
\bibitem [{\citenamefont {Hadjiconstantinou}\ \emph {et~al.}(2003)\citenamefont
  {Hadjiconstantinou}, \citenamefont {Garcia}, \citenamefont {Bazant},\ and\
  \citenamefont {He}}]{jcperror}%
  \BibitemOpen
  \bibfield  {author} {\bibinfo {author} {\bibfnamefont {Nicolas~G.}\
  \bibnamefont {Hadjiconstantinou}}, \bibinfo {author} {\bibfnamefont
  {Alejandro~L.}\ \bibnamefont {Garcia}}, \bibinfo {author} {\bibfnamefont
  {Martin~Z.}\ \bibnamefont {Bazant}}, \ and\ \bibinfo {author} {\bibfnamefont
  {Gang}\ \bibnamefont {He}},\ }\bibfield  {title} {\enquote {\bibinfo {title}
  {Statistical error in particle simulations of hydrodynamic phenomena},}\
  }\href {\doibase https://doi.org/10.1016/S0021-9991(03)00099-8} {\bibfield
  {journal} {\bibinfo  {journal} {Journal of Computational Physics}\ }\textbf
  {\bibinfo {volume} {187}},\ \bibinfo {pages} {274 -- 297} (\bibinfo {year}
  {2003})}\BibitemShut {NoStop}%
\bibitem [{\citenamefont {Berendsen}\ \emph {et~al.}(1984)\citenamefont
  {Berendsen}, \citenamefont {Postma}, \citenamefont {van Gunsteren},
  \citenamefont {DiNola},\ and\ \citenamefont {Haak}}]{Berendsen}%
  \BibitemOpen
  \bibfield  {author} {\bibinfo {author} {\bibfnamefont {H.~J.~C.}\
  \bibnamefont {Berendsen}}, \bibinfo {author} {\bibfnamefont {J.~P.~M.}\
  \bibnamefont {Postma}}, \bibinfo {author} {\bibfnamefont {W.~F.}\
  \bibnamefont {van Gunsteren}}, \bibinfo {author} {\bibfnamefont
  {A.}~\bibnamefont {DiNola}}, \ and\ \bibinfo {author} {\bibfnamefont {J.~R.}\
  \bibnamefont {Haak}},\ }\bibfield  {title} {\enquote {\bibinfo {title}
  {Molecular dynamics with coupling to an external bath},}\ }\href {\doibase
  http://dx.doi.org/10.1063/1.448118} {\bibfield  {journal} {\bibinfo
  {journal} {J. Chem. Phys.}\ }\textbf {\bibinfo {volume} {81}},\ \bibinfo
  {pages} {3684--3690} (\bibinfo {year} {1984})}\BibitemShut {NoStop}%
\end{thebibliography}%
\bibstyle{is-unsrt}
\end{document}